\documentclass[aps,twocolumn,showpacs,preprintnumbers]{revtex4}

\usepackage{graphicx}
\usepackage{dcolumn}
\usepackage{bm}
\usepackage{subfigure}

\newcommand{\beq}{\begin{eqnarray}}
  \newcommand{\eeq}{\end{eqnarray}}

\begin{document}



\title{Oscillations of a solid sphere falling through a wormlike micellar 
fluid}

\author{Anandhan Jayaraman and Andrew Belmonte}

\address{\it The W.~G.~Pritchard Laboratories, 
Department of Mathematics,\\
Pennsylvania State University, University Park, PA 16802}

\date{\today}


\begin{abstract}
We present an experimental study of the motion of a solid sphere 
falling through a wormlike micellar fluid.  While smaller or lighter 
spheres quickly reach a terminal velocity, larger or heavier spheres 
are found to oscillate in the direction of their falling motion.  The 
onset of this instability correlates with a critical value of the 
velocity gradient scale $\Gamma_{c}\sim 1$ s$^{-1}$.  We relate this 
condition to the known complex rheology of wormlike micellar fluids, 
and suggest that the unsteady motion of the sphere is caused by the 
formation and breaking of flow-induced structures.
\end{abstract}


\maketitle



\begin{figure}[h]                             
\includegraphics[height=2in]{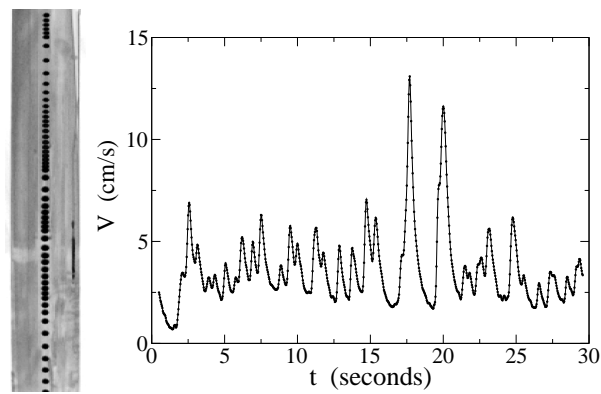} 
\includegraphics[height=1.9in]{NewTef1-4c.eps} 

\caption{Collage of video images showing the descent of a $3/16''$ 
diameter teflon sphere in an aqueous solution of 6.0~mM  CTAB/NaSal
(image shown is 50 cm in height, with $\Delta t$ = 0.13 s); b) velocity
vs.~time for a $1/4''$  diameter teflon sphere falling through  9.0 mM 
CTAB/NaSal.}

\label{f-1}
\end{figure}

A sphere falling through a viscous Newtonian fluid is a classic 
problem in fluid dynamics, first solved mathematically by Stokes in 
1851 \cite{stokes51}.  Stokes provided a formula for the drag force 
$F$ experienced by a sphere of radius $R$ when moving at constant 
speed $V_{0}$ though a fluid with viscosity $\mu$: $F= 6\pi\mu R 
V_{0}$.  The simplicity of the falling sphere experiment has meant 
that the viscosity can be measured directly from the terminal velocity 
$V_{0}$, using a modified Stokes drag which takes into account wall 
effects \cite{clift78}.  The falling sphere experiment has also been 
used to study the viscoelastic properties of many polymeric 
(non-Newtonian) fluids \cite{ballvisc84,chhabra93,bird87}.  In 
general, a falling sphere in a non-Newtonian fluid always approaches a 
terminal velocity, though sometimes with an oscillating transient 
\cite{walt92,raja96,arigo97}.  In this paper we present evidence that 
a sphere falling in a wormlike micellar solution does not seem to 
approach a steady terminal velocity; instead it undergoes continual 
oscillations as it falls, as shown in Fig.~\ref{f-1}.

A wormlike micellar fluid is an aqueous solution in which amphiphilic 
(surfactant) molecules self-assemble in the presence of NaSal into 
long tubelike structures, or worms \cite{israel}; these 
micelles can sometimes be as long as 1$\mu m$ \cite{larsonBK}.  Most 
wormlike micellar solutions are viscoelastic, and at low shear rates 
their rheological behavior is very similar to that of polymer 
solutions.  However, unlike polymers, which are held together by 
strong covalent bonds, the micelles are held together by relatively 
weak entropic and screened electrostatic forces, and hence can break 
under shear.  In fact, under equilibrium conditions these micelles are 
constantly breaking and reforming, providing a new mechanism for 
stress relaxation \cite{cates90}.

The nonlinear rheology of these micellar fluids can be very different 
from standard polymer solutions \cite{cates90,spenley93,walk01}.  
Several observations of new phenomena have been reported, including 
shear thickening \cite{liu96,hu98}, a stress plateau in steady shear 
rheology \cite{cappel97,porte97}, and flow instabilities such as 
shear-banding \cite{lerouge00}.  Bandyopadhyay {\it et al.} have 
observed chaotic fluctuations in the stress when a wormlike micellar 
solution is subjected to a step shear rate above a certain critical 
value (in the plateau region of stress-shear rate curve) \cite{chaos}.  
A shear-induced transition from an isotropic to a nematic micellar 
ordering has also been observed \cite{berret94}.

There is increasing experimental evidence relating the onset of some 
of these interesting rheological phenomena to the formation of 
mesoscale aggregations, the so-called ``shear-induced structures'' 
(SIS) \cite{rehage82,walk01}.  These structures have now been imaged 
directly in wormlike micellar fluids using electron microscopy 
\cite{keller98}.  Recent experiments have correlated the formation of 
SIS with the occurence of shear thickening \cite{liu96} and shear 
banding \cite{call01}.  Also, visual observations have shown that the 
growth of SIS is followed by their tearing or breaking, after which 
they grow again \cite{liu96,wheel98}.

Recently, oscillations in the shape and velocity of a rising bubble 
have been observed in the wormlike micellar system 
cetyltrimethylammonium bromide (CTAB)~/ sodium salicylate (NaSal), for 
concentrations from 8 - 11 mM \cite{ab00}.  These oscillations occur 
when the bubble volume is greater than a certain critical value; small 
bubbles, which remain spherical or ellipsoidal, do not oscillate.  The 
bubble develops a cusp as it rises through the solution.  At the 
moment of cusp formation, the bubble suddenly ``jumps'' releasing the 
cusp.  A strong negative wake \cite{hass79} is observed behind the 
bubble after every jump.  Similar dynamics are also observed in other 
wormlike micellar systems \cite{hand02}.

The fact that shape oscillations are coupled with the velocity 
oscillations of the bubble suggests that the oscillations are caused 
by surface tension effects.  The observations presented here, that a 
solid sphere also oscillates while falling through a CTAB/NaSal 
solution, indicate that such oscillations are not due to surface 
tension or cusp-like tails \cite{ab00}.  We conjecture that the 
oscillations are due to the formation of flow-induced structures in 
the combined shear and extensional flow around the sphere.


{\bf Experimental Setup and Results.}
Our study focuses on the micellar system CTAB/NaSal 
\cite{shik88,liu96,cappel97}, one of several aqueous solutions 
containing the organic salt sodium salicylate (NaSal), which 
facilitates the formation of long tubular ``wormlike'' micelles of 
cationic surfactants \cite{rehage91}.  The CTAB and NaSal used here 
are obtained from Aldrich, and dissolved in distilled deionized water 
without further purification.  The fluids are mixed for several days, 
then allowed to settle for a day before use.  We restrict our 
solutions to the molar ratio 1:1 \cite{shik88}, and have found that 
only for a range of concentrations around 10 mM can we obtain a low 
Reynolds number flow which is still experimentally convenient to use.  
All data presented here are for 9.0 mM equimolar CTAB/NaSal solutions
(except for Fig.~1a, which is for 6.0 mM CTAB/NaSal).  
At this concentration, the solution is known to form wormlike micelles 
\cite{liu96}.

\begin{figure}                                            
\includegraphics[width=3in]{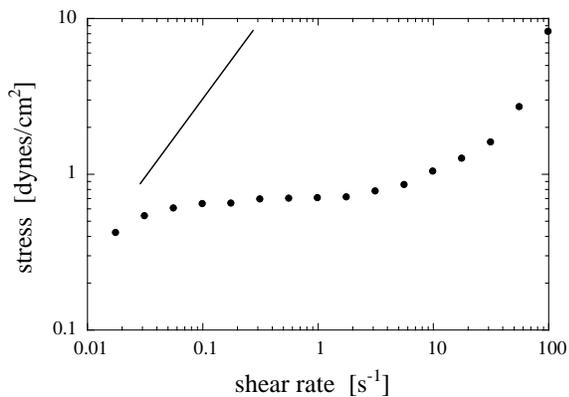} 

\caption{Shear rheology of the 9mM CTAB/NaSal wormlike micellar 
solution: stress vs.~shear rate data obtained using a Couette 
rheometer at 30$^{\circ}$ C. For comparison, the straight line 
indicates the scaling of a Newtonian fluid (linear).}

\label{f-rheo}
\end{figure}

The shear rheology of our fluid was investigated using a concentric 
Couette rheometer (Rheometrics RFS III).  The steady state stress is 
shown as a function of applied shear rate in Fig.~\ref{f-rheo}.  On 
this log-log plot, a linear slope is superimposed for comparison - the 
fluid is clearly non-Newtonian in shear \cite{bird87}.  Additionally, 
the fluid shows a plateau in the stress over a range of shear rates 
(approximately from 0.2 to 2 s$^{-1}$), a characteristic of wormlike 
micellar fluids \cite{larsonBK,berret97}.

Our experimental setup consists of a tall cylindrical cell with inner 
diameter $D= 9$ cm and length $L =120 $ cm, filled with the 
experimental fluid.  For all of our spheres $d/D \leq 0.21$, where $d$ 
is the diameter of the sphere.  The cell is enclosed in a 120 cm high 
box, and recirculating water in this surrounding volume is used to 
control the temperature of the fluid.  Our experiments are performed 
at $T = 30.0^{\circ}$ C. The spheres (made of nylon, delrin, or 
teflon) are dropped in the center of the tube using tweezers, and 
allowed to fall through 10 cm before data is taken.  The fluid was 
allowed to relax for at least 60 minutes between each drop.  The 
motion of the sphere is captured by a CCD camera and the movie is 
stored digitally in a computer.  An in-house image analysis program 
is then used to extract the velocity vs time data of the falling 
sphere from the digital images.


\begin{figure}
\centering
{
 \includegraphics[width=4cm]{del1_4c.eps}
 }
\hspace{0.1cm}
{
\includegraphics[width=4cm]{del3_8b.eps}
}
\label{fig:sub} 
\end{figure}


\begin{figure}
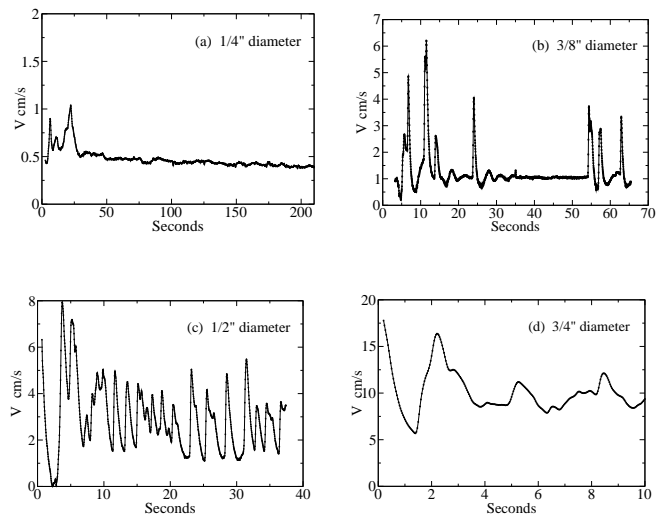

\centering
{
 \includegraphics[width=4cm]{del1_2b.eps}
 }
\hspace{0.1cm}
{
\includegraphics[width=4cm]{del3_4aa.eps}
}
\caption{The velocity of delrin spheres with diameters of a) $1/4''$, b) 
$3/8''$, c) $1/2''$, and d) $3/4''$.}
\label{f-2} 
\end{figure}


While small spheres reach a terminal velocity after some transient 
oscillations (Fig.~\ref{f-2}a), spheres of larger size do not seem to 
approach a terminal velocity - they oscillate as they fall 
(Fig.~\ref{f-2}b-d).  These oscillations are not perfectly periodic, 
displaying some irregularity.  However, an average frequency can be 
defined as the number of oscillations over the entire fall divided by 
the time taken.  The average frequency of oscillations increases with 
the radius or volume of the sphere, as was also observed for the 
oscillations of a rising bubble \cite{ab00}.  Although the amplitude 
of these velocity oscillations vary widely for a given sphere, the 
oscillations show a common characteristic of a sudden acceleration and 
a relatively slower deceleration.  At the moment of each acceleration, 
a strong negative wake \cite{hass79} is visually observed as a recoil 
in the fluid, as if the fluid were letting go of the sphere.

\begin{figure}
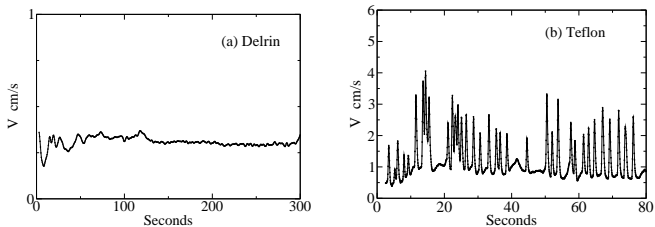

\centering
{
 \label{fig:sub:a}
 \includegraphics[width=4.05cm]{del3_16ac.eps}
 }
 \hspace{0.07cm}
{
\label{fig:sub:b}
\includegraphics[width=4.05cm]{tef3_16b.eps}
}
\caption{The velocity of two 3/16'' diameter spheres
with a) $\rho =1.35$~g/cm$^3$ (delrin),  and b) $\rho =2.17$~g/cm$^3$ (teflon). The lighter
sphere reaches a terminal velocity, whereas the heavier sphere
exhibits the unsteady behavior.}
\label{f-3} 
\end{figure}

For a sphere of fixed radius, there is a transition to nontransient 
oscillations as the density of the sphere is increased.  As seen in 
Fig.~\ref{f-3}, a delrin sphere (density $\rho =1.35$~g/cm$^3$) with 
$d = 3/16''$ does not oscillate, however a teflon sphere ($\rho 
=2.17$~g/cm$^3$) of the same diameter does oscillate.  If however the 
sphere is very large or heavy, it falls through the fluid very fast, 
with less well-defined oscillations (Fig.~\ref{f-2}d).

To characterize the onset of these oscillations, we estimate the 
average magnitude of the velocity gradients $\Gamma$ using the ratio 
of the velocity of the sphere to its diameter: $\Gamma = V_{b}/d$, 
where we take $V_{b}$ to be the baseline velocity below the 
oscillations.  For the four experiments shown in Fig.~\ref{f-2}, we 
find $\Gamma \simeq$ 0.79, 1.05, 1.2, and 4.2 s$^{-1}$, 
respectively.  For Fig.~\ref{f-3}, we find $\Gamma \simeq$ 0.63, 
and 1.7 s$^{-1}$, respectively.  Thus the oscillations appear to start 
at a critical value $\Gamma_{c} \sim 1$ s$^{-1}$.  Note that this 
average velocity gradient corresponds to a shear rate lying in the 
`flat' region of the shear stress curve (Fig.~\ref{f-rheo}).


{\bf Discussion.} The first observation of nontransient oscillations 
of an object in a wormlike micellar fluid was made for a rising bubble 
\cite{ab00}.  The fact that similar oscillations are observed for a 
falling sphere indicates that the oscillations are not caused by a 
surface instability.  The sudden acceleration of the sphere during the 
oscillations indicates that the drag on the sphere has suddenly 
decreased.  We hypothesize that this sudden drop in the drag is due to 
the breakup of flow-induced structures (FIS) that are formed in the 
region around the sphere.

Using small angle light scattering on a CTAB/NaSal micellar solution 
in a Couette cell, Liu \& Pine found that once the fluid was subjected 
to shear rates greater than a certain critical value, streaks were 
observed in the scattering intensities, indicating the formation of 
mesoscale structures \cite{liu96}.  These elongated structures have a 
width $\sim 1\mu$m, which is much larger than the diameter of a single 
micelle (5 nm) \cite{keller98}.  The appearance of these structures 
dramatically increased the apparent viscosity of the fluid 
\cite{call01}, and also made the fluid very elastic.  Moreover, the 
structures do not appear instantaneously when the fluid is sheared - 
they form only after a finite induction time ($\sim$ seconds).  They 
grow from the stationary cylinder to the moving cylinder in the 
Couette cell, but are ripped apart as they approach the moving 
surface.  The cycle begins again as the structures start growing from 
the stationary cylinder.  The process of growing and ripping can be 
correlated to macroscopic stress fluctuations.  These experiments 
suggest that FIS cannot sustain large 
stresses.  Although these observations were made for a low 
concentration shear thickening CTAB/NaSal solution, the FIS are 
believed to form even for higher concentrations; similar observations 
of birefringent structures and stress oscillations were made by 
Wheeler {\it et al.}~in 40~mM CPCl/NaSal \cite{wheel98}.

We propose the following picture as a mechanism for the oscillations: 
if a sphere falls fast enough through a wormlike micellar fluid, FIS 
are formed around the sphere which increase the effective viscosity, 
and thus the drag.  However, when the stress around the sphere reaches 
a large enough value, these structures break and the fluid in the wake 
of the sphere recoils.  This causes a sudden drop in the drag 
experienced by the sphere, which suddenly accelerates.  Once the 
sphere moves into fresh fluid, FIS start forming again, repeating the 
cycle.

The apparent critical velocity gradient $\Gamma_{c}$ required for the 
onset of oscillations of a falling sphere indicates a critical 
gradient required for the formation of the FIS. Since the formation of 
these structures is not a regular, periodic process 
\cite{liu96,wheel98}, there are fluctuations in the drag experienced 
by the sphere, which would cause irregular oscillations.  This would 
also explain why small spheres (and small bubbles) do not oscillate: 
the lower terminal velocity does not shear or stretch the fluid enough 
to produce the structures.  Effectively, lighter spheres move in a 
more uniform fluid than spheres of moderate weight.

On the other hand, very heavy spheres do not seem to have well defined 
oscillations (see Fig.~\ref{f-2}d).  This could be an effect of the 
inertia of the sphere.  It could also be due to the time needed to 
form shear-induced structures \cite{hu98,wheel98}.  If the sphere is 
heavy its average velocity is sufficiently high and it moves into 
fresh fluid while the structures are still forming.

According to our interpretation, the large negative wake behind the 
sphere seen after every ``jump'' would be due to the shedding of these 
FIS. The onset of oscillations in the bubbles at the formation of a 
cusp indicates that the surrounding fluid is unable to support the 
high stresses required to form the cusp.  When the flow around the 
bubble induces the bubble to form a cusp, the stress becomes large 
enough to break the FIS, which are subsequently shed.  The shedding 
is seen as a strong negative wake behind the bubble \cite{ab00}.  This 
explanation also enables us to make a prediction: every cusped bubble 
in a micellar fluid which allows for FIS should oscillate.  
Experiments with bubbles in our CTAB/NaSal solution are consistent 
with this prediction.



{\bf Conclusions.} A sphere falling in a viscous Newtonian fluid 
reaches a steady terminal velocity; the approach to this terminal 
velocity can be shown to be monotone \cite{ejde}, in agreement with 
observations.  In polymeric fluids, a final steady state is also 
always observed experimentally.  In this paper we have shown that the 
unusual behavior seen in the flow of wormlike micellar solutions 
extends to the classic problem of flow past a sphere.

We believe that the nontransient oscillations of the falling sphere 
are caused by the formation of flow-induced structures.  While the 
rheology of wormlike micellar fluids has typically focused on either 
pure shear flow \cite{rehage91,berret97} or pure extensional flow 
\cite{walk96}, the velocity field produced by a falling sphere is more 
complicated: the fluid near the surface is being sheared, whereas in 
the wake the fluid is primarily under extensional flow.  It is thus 
not what is called a {\it viscometric flow} \cite{bird87}.  Recently 
some indirect evidence has been seen of structures produced by 
extensional flows \cite{smolka}.  The observations presented here may 
be an example of the dynamics of these flow-induced structures in a 
complicated hydrodynamic flow.

Analyzing this problem certainly presents a challenge, and 
mathematical modeling will probably require full scale numerical 
simulations.  Numerical studies of the transient motion of a falling 
sphere have only recently been undertaken even for polymer fluids (see 
e.g.~\cite{raja96} and references therein).  The constitutive equation 
chosen to model our experiment should give a good fit to both the 
shear stress and the extensional flow data, and in addition include 
the as yet unknown property which produces nontransient oscillations.

The shear-stress flow curve for our wormlike micellar fluid displays a 
flat region (see Fig.~\ref{f-rheo}), which is believed to be a 
manifestation of a theoretical non-monotone stress-shear rate relation 
\cite{spenley93,lu00}.  It is well known that steady-shear flow in the 
decreasing region of a non-monotonic flow curve is unstable (it is 
ill-posed in the Hadamard sense \cite{joseph}).  Such an instability 
has been attributed to cause different physical effects such as 
shear-banding in micellar solutions \cite{BSPP}, and the shark-skin 
\cite{joseph} and spurt \cite{yy95} instabilities in polymer melts.  
Heuristically, the onset of oscillations of a falling sphere could be 
due to the same instability.  The fact that the average shear rate at 
onset of oscillations sits in the flat region of the shear stress 
curve supports this conjecture.  To test this further one should 
choose a constitutive equation which displays a non-monotonic flow 
curve (for instance the Johnson-Segalman equation \cite{js77,lu00}) 
and simulate the falling sphere problem.  Such a simulation is 
currently in progress.


We thank Lynn Walker, H. A. Stone, N. Handzy, J. Jacobsen and Jinchao 
Xu for discussions.  AB acknowledges the support of the Alfred P. 
Sloan Foundation, and National Science Foundation (CAREER Award 
DMR-0094167).


\end{document}